\documentclass[journal,onecolumn]{IEEEtran}

\usepackage{amssymb}
\usepackage{graphicx}
\usepackage{subfig}

\usepackage{listings}
\usepackage{balance}

\usepackage{cite}

\usepackage[cmex10]{amsmath}
\usepackage{setspace}
\usepackage[cmex10]{amsmath}

\usepackage[acronym]{glossaries}

\makeglossaries

    \newglossaryentry{Crosscutting}{name={crosscutting}, description={Properties or areas of interest such as quality of service, energy consumption, location awareness, users' preferences, and security}}
\protect    \newglossaryentry{heterogeneous}{name={heterogeneous}, description={A set of collaborated aspects (code fragments), that extend the application behaviour in several parts of the program and have an impact across the whole software system}}
\protect    \newglossaryentry{Homogeneous}{name={homogeneous}, description={Applying the same code, that extend the application behaviour in several parts of the program}}

    \newglossaryentry{COCA-platform}{name={COCA-platform}, description={Context-Oriented Component-based Applications and the middleware architecture}}
   \newglossaryentry{COCA-middleware}{name={COCA-middleware}, description={Context-Oriented Component-based Applications Middleware}}
 \newglossaryentry{COCA-component}{name={COCA-component}, description={Context-Oriented Component-based Applications Component}}

 \newglossaryentry{COCA-ADL}{name={COCA-ADL}, description={Context-Oriented Component-based Applications Architecture Description Language}}
\protect  \newglossaryentry{Context}{name={context}, description={Any information that is computationally accessible and upon which behavioural variations depend}}
\protect    \newglossaryentry{Self-adaptive}{name={self-adaptive}, description={A self-adaptive application modifies its own structure and behaviour in response to changes in its operating environment}}
   
 \protect    \newglossaryentry{Self-* properties}{name={Self-* properties}, description={The autonomic properties of a software, which includes (self-organising, self-healing, self-optimising and self- protecting)}}
 \protect   \newglossaryentry{Self-organising}{name={Self-organising}, description={The capability of reconfiguring automatically and dynamically in response to changes by installing, updating, integrating, and composing/decomposing software entities}}
 \protect   \newglossaryentry{Self-healing}{name={Self-healing}, description={The capability of discovering, diagnosing and reacting to disruptions. It can also anticipate potential problems, and accordingly take proper actions to prevent a failure}}
\protect \newglossaryentry{Self-optimising}{name={Self-optimising}, description={The capability of managing performance and resource allocation in order to satisfy the requirements of different users. End-to-end response time, throughput, utilisation and workload are examples of important concerns related to this property}}   
 \protect \newglossaryentry{Self-protecting}{name={Self-protecting}, description={The capability of detecting security breaches, anticipating problems and recovering from their effects. It has two aspects, namely defending the system against malicious attacks, and anticipating problems and taking actions to avoid them or mitigate their effects}}  
\protect   \newglossaryentry{Context-dependent}{name={context-dependent}, description={A context-dependent application adjusts its behaviour according to context conditions arising during execution}} 
    
    \newglossaryentry{Component Collaboration Architecture}{name={Component Collaboration Architecture}, description={how to model the structure and the behaviour of components at varying and mixed levels of granularity}} 
     
  \protect       \newglossaryentry{AspectJ}{name={AspectJ}, description={A Java implementation of Aspect-Oriented Programming paradigm}} 

\newglossaryentry{ECORE}{name={ECORE}, description={A meta model in the Eclipse Modelling Framework}} 
\newglossaryentry{Context-awareness}{name={context-awareness}, description={The software system is aware of its context, which is its operational environment}} 

\protect \newglossaryentry{Context-aware}{name={context-aware application}, description={refer to a class of software systems that are able to monitor and detect context changes in the environment where they operate}} 
 \newglossaryentry{Theme/UML}{name={Theme/UML}, description={The Theme Approach is an aspect-oriented methodology that encompasses the requirements analysis, design and mapping to implementation phases of the development life cycle}} 
 
\newglossaryentry{MUSIC}{name={MUSIC}, description={An open source platform for the development and execution of self-adapting, reconfigurable mobile applications which are dynamically adapted to the user and execution context}} 
\newglossaryentry{U-MUSIC}{name={U-MUSIC}, description={A development methodology that  extends the works from the MADAM and the MUSIC platforms in order to support unanticipated adaptation
}} 

\newglossaryentry{OSGI-framework}{name={OSGI-framework}, description={A module system and service platform for the Java programming language that implements a complete and dynamic component model}} 
\newglossaryentry{CAUCE}{name={CAUCE}, description={Model-driven Development of Context-aware Applications for Ubiquitous Computing Environments}} 

\newglossaryentry{DESMET}{name={DESMET}, description={A methodology for evaluating software engineering methods and tools}}

\newglossaryentry{JCOOL}{name={JCOOL}, description={A COntext Oriented Language tailored to handle context awareness in Java applications}} 
 
\newglossaryentry{JCOP}{name={JCOP}, description={A context-oriented programming (COP) extension to the Java programming language}} 

\newglossaryentry{MAPE-K}{name={MAPE-K}, description={The autonomic computing closed loop, which includes Monitoring, Analysing, Planning and Executing}} 
 \protect \newglossaryentry {Joinpoint}{name={joinpoint}, description={ A point in the control flow of a program. In aspect-oriented programming a set of join points is described as a pointcut. A join point is a specification of when, in the corresponding main program, the aspect code should be executed.}} 

\protect \newglossaryentry {Aspect}{name={aspect}, description={An aspect of a program is a feature linked to many other parts of the program, but which is not related to the program's primary function. An aspect crosscuts the program's core concerns, therefore violating its separation of concerns that tries to encapsulate unrelated functions}} 
\protect \newglossaryentry {Pointcut}{name={pointcut}, description={ Is a set of join points. Whenever the program execution reaches one of the join points described in the pointcut, a piece of code associated with the pointcut (called advice) is executed. This allows a programmer to describe where and when additional code should be executed in addition to an already defined behaviour.}} 
  \newglossaryentry {Advice}{name={advice}, description={Describes a class of functions which modify other functions when the latter are run; it is a certain function, method or procedure that is to be applied at a given join point of a program.}} 
 
 \protect \newglossaryentry {Separation of concerns}{name={separation of concerns}, description={Is the process of separating a computer program into distinct features that overlap in functionality as little as possible.}} 

\newacronym[\glsshortpluralkey=cas,\glslongpluralkey=contrived acronyms]{aca}{aca}{a contrived acronym}
\newacronym{cosd}{COSD}{Context Oriented Software Development}
\newacronym{MDA}{MDA}{Model Driven Architecture}
\newacronym{COP}{COP}{Context-Oriented Programming}
\newacronym{AOP}{AOP}{Aspect-Oriented Programming}
\newacronym{DAOP}{DAOP}{Dynamic Aspect Oriented Programming}
\newacronym{AOSD}{AOSD}{ Aspect Oriented Software Development}
\newacronym{MDD}{MDD}{ Model Driven Development}
\newacronym{CBSD}{CBSD}{ Component-based Software Development}

\newacronym{ADL}{ADL}{ Architecture Description language}
 \newacronym{COCOMO II}{COCOMO II}{ The Constructive Cost Model II}
  
 \newacronym{PIV}{PIV}{ Platform Independent View}
 \newacronym{CIV}{CIV}{ Computation Independent View}
 \newacronym{PSV}{PSV}{ Platform Specific View}
 
\newacronym{PIM}{PIM}{ Platform Independent Model}
 \newacronym{CIM}{CIM}{Computation Independent Model}
 \newacronym{PSM}{PSM}{Platform Specific Model}
   \newacronym{QoS}{QoS}{Quality of Services}
    \newacronym{OOP}{OOP}{Object Oriented Programming}

   \newacronym{ATAM}{ATAM}{Architecture Trade-off Analysis Method }
   \newacronym{UML}{UML}{Unified Modelling Language }
    \newacronym{OMG}{OMG}{Object Management Group}
     \newacronym{ECA}{ECA}{ Enterprise Collaboration Architecture}
      \newacronym{EDOC}{EDOC}{The Enterprise Distributed Object Computing}
 
      \newacronym{CCA}{CCA}{Component Collaboration Architecture}
       \newacronym{MOF}{MOF}{Meta Object Facility }
       \newacronym{DSL}{DSL}{Domain Specific Language}
       \newacronym{ATL}{ATL}{Atlas Transformation Language}
       \newacronym{EMF}{EMF}{Eclipse Modelling Framework}
       
              \newacronym{MADAM}{MADAM}{Mobility and ADaptation enAbling Middleware}
              \newacronym{OSGI}{OSGI}{Open Services Gateway initiative framework}
              
                            \newacronym{MOSEL}{MOSEL}{modelling Specification and Evaluation Language}
                            \newacronym{PM}{PM}{Person-Months}  
                            \newacronym{TDEV}{TDEV}{Time to Develop}
                            \newacronym{CAMEL}{CAMEL}{Context	Awareness modelling Language}

                            \newacronym{A-MUSE}{A-MUSE}{Architectural modelling for Service Enabling in Freeband}

                            \newacronym{CASA}{CASA}{Contract-based Adaptive Software Architecture}

                            \newacronym{POI}{POI}{Places Of Interest}
                            \newacronym{SLOC}{SLOC}{Source Lines Of Code}
 
                            \newacronym{ASLOC}{ASLOC}{Adapted source lines of code}
                             \newacronym{UFP}{UFP}{Unadjusted Function Points}
                             \newacronym{RFID}{RFID}{Radio Frequency IDentification}
                              \newacronym{QR-code}{QR-code}{Quick Response Code}
                                \newacronym{DP}{DP}{Decision Point}
                                \newacronym{DPL}{DPL}{Decision PoLicy}
                                \newacronym{CCTV}{CCTV}{Closed-Circuit TeleVision}
                                 \newacronym{COCA-MDA}{COCA-MDA}{Context-Oriented Component-based Applications Model-Driven Architecture}

\begin{document}

\title{A Framework for Evaluating Model-Driven Self-adaptive Software Systems}
 

\author{\IEEEauthorblockN{Basel Magableh}
\IEEEauthorblockA{\textit{School of Computer Science, } \\
\textit{Dublin Institute of Technology,}\\
\textit{Technological University}\\
Dublin, Ireland \\
basel.magableh@dit.ie}}



\maketitle
\begin{abstract}

In the last few years, \gls{MDD}, \gls{CBSD}, and context-oriented software have become interesting alternatives for the design and construction of self-adaptive software systems. In general, the ultimate goal of these technologies is to be able to reduce development costs and effort, while improving the modularity, flexibility, adaptability, and reliability of software systems. An analysis of these technologies shows them all to include the principle of the \gls{Separation of concerns}, and their further integration is a key factor to obtaining high-quality and self-adaptable software systems. Each technology identifies different concerns and deals with them separately in order to specify the design of the self-adaptive applications, and, at the same time, support software with adaptability and context-awareness. This research studies the development methodologies that employ the principles of model-driven development in building self-adaptive software systems. To this aim, this article proposes an evaluation framework for analysing and evaluating the features of model-driven approaches and their ability to support software with self-adaptability and dependability in highly dynamic contextual environment.  Such evaluation framework can facilitate the software developers on selecting a development methodology that suits their software requirements and reduces the development effort of building self-adaptive software systems. This study highlights the major drawbacks of the propped model-driven approaches in the related works, and emphasise on considering the volatile aspects of self-adaptive software in the analysis, design and implementation phases of the development methodologies. In addition, we argue that the development methodologies should leave the selection of modelling languages and modelling tools to the software developers.
  \end{abstract}

\section{Introduction}

 There is a growing demand for developing applications with aspects such as \gls{Context} awareness and self-adaptive behaviours. \gls{Context} awareness \cite{Parashar:2005p4081} means that the system is aware of its context, which is its operational environment. Hirschfeld et al. \cite{Hirschfeld:2008p1620} considered context to be any information that is computationally accessible and upon which behavioural variations depend. A self-adaptive application adjusts its behaviour according to context conditions arising during execution. A self-adaptive application modifies its own structure and behaviour in response to changes in its operating environment \cite{Oreizy:1999p3722}. 

In recent years, a significant number of model-driven architecture approaches were proposed for the construction of context-dependent and self-adaptive applications. The \gls{OMG} presented \gls{MDA} as a set of guidelines for building software systems based on the use of the \gls{MDD} methodology \cite{OMG:2010p3514}. MDA focuses primarily on the functionality and behaviour of a distributed application or system deployed across many platforms. In MDA, the functionality and behaviour are modelled once and only once. Thus, MDA defines the notions of a \gls{CIM}, \gls{PIM} and \gls{PSM}. \gls{CIM} describes the software requirements in computational free fashion. A \gls{PIM} describes the parts of a solution that do not change from one platform to another, and a \gls{PSM} includes descriptions of parts that are platform dependent \cite{OMG:2010p3514}. 

This article contributes to the knowledge by providing an evaluation framework for most popular model-driven approaches that were proposed to support the development of self-adaptive software systems. Such evaluation framework can facilitate software developers on selecting the best development approach that suits their needs and the software requirements.  
 This article is structured as follows: the self-adaptive software and their self-* properties are defined in Section \ref{sec:selfadaptation}. Section \ref{sec:modelling} provides a detailed description of model-driven approaches that were proposed for facilitating the development of self-adaptive software systems. Section \ref{sec:feature} proposes an evaluation framework for  analysing and evaluating those model-driven approaches. Section \ref{sec:res} illustrates the evaluation results, followed by the conclusions of this study. 

\section{Self-adaptive Software\label{sec:selfadaptation}}

Mobile computing infrastructures make it possible for mobile users to run software systems in heterogeneous and resource-constrained platforms. Mobility, heterogeneity, and device limitations create a challenge for the development and deployment of mobile software. Mobility induces \gls{Context} changes to the computational environment and therefore changes to the availability of resources and services. This requires software systems to be able to adapt their functionality/behaviour to the context changes \cite{Inverardi:2009p2345}. This class of software systems are called \gls{Context-dependent}/self-adaptive applications, which have the ability to modify their own structure and behaviour in response to context changes in the environment where they operate \cite{Oreizy:1999p3722}.  Self-adaptive software offers the users with context-dependent and context-independent functionality. Context-independent functionality (also called base functionality) refers to software functionality whose implementation is unaffected by the context changes. For example, the map view and user login forms in mobile map application are context-free functionality (i.e. context changes would not change their functionality). The context-dependent functionality refers to software functionality, which exhibits volatile behaviour when the context changes. Self-adaptive software can be seen as a collaboration of individual features spanning the software modules in several places \cite{Hirschfeld:2008p1620}, and they are sufficient to qualify as \gls{heterogeneous} crosscutting in the sense that different code fragments are applied to different program parts \cite{Apel:2006p3702}. Before encapsulating crosscutting context-dependent behaviours into a software module, the developers must first identify the behaviours in the software requirements. This is difficult to achieve because, by their nature, context-dependent behaviours are entangled with other behaviours, and are likely to be included in multiple parts (scattered) of the software modules \cite{Lincke:2010p4207}. Using intuition or even domain knowledge is not necessarily sufficient for identifying their volatile behaviour; instead, a formal procedure is needed for analysing and separating their individual concerns \cite{Carton:2007p1466}.

Implementing a self-adaptive software system in a resource-poor environment faces a wide rang of variance in platforms' specifications and \gls{QoS} \cite{Kuwadekar:5558237}. Mobile devices have different capabilities in terms of CPU, memory, and network bandwidth \cite{khaled:2008p189278}. Everything from the devices used and resources available to network bandwidths and user context can change extremely at runtime \cite{Belaramani:2003p3679}. 

In general, model-driven architecture has several challenges such as maintaining the correspondence between architectural models and software implementation in order to ensure that architecture-based adaptation is appropriately executed. The second issue is providing the necessary facilities for implementing the software in wide-range of platforms. An appropriate way to study those challenges is to classify them on the basis of adaptation features that they support and how they manage software variability in the model level \cite{Salehie:2009p3693}. To this aim, the following sections describe several model-driven approaches proposed for facilitating the engineering self-adaptive software system. 

\section{Modelling Self-adaptive Software \label{sec:modelling}}
 In the classical view of object-oriented software development, the modular structure for software systems has rested on several assumptions. These assumptions may no longer characterize the challenge of constructing self-adaptive software systems that are to be executed in mobile computing environments \cite{Harrison:2011960313}. The most important assumptions in object-oriented development methodologies are that the decision to use or reuse a particular component/object is made at the time the software is developed. However, the development of a variety of modern self-adaptive software architectures such as mobile/ubiquitous computing, and component-based and context-oriented software has emphasized on deferring these decisions about component selection until runtime. This might increase the software capabilities in terms of variability, adaptability, and maintainability, and increase the anticipatory level of the software by loading a particular component/service that can handle unforeseen context changes dynamically. 
 
Supporting the development and execution of self-adaptive software systems raises numerous challenges. These challenges include: 1) the development processes for building them, 2) the design space, which describes the design patterns and the best practices of designing their building blocks, i.e. component model or code fragments, 3) the adaptation mechanism that describes the best adaptation action that can be used under the limited resources of the computational environment.

In general, the development self-adaptive systems faces several challenges such as maintaining the correspondence between architectural models and system implementation in order to ensure the adaptation action is appropriately executed. The second issue is providing the necessary configuration that suits the deployment platforms. An appropriate way to study the challenges is to classify them on the basis of adaptation features that they support and how they manage software variability in the architecture level \cite{Salehie:2009p3693}. In the following sections, several approaches that target engineering self-adaptive systems are discussed. 
\subsection{Model Driven Development and AOP}

Carton et al. \cite{Carton:2007p1466} proposed the Theme/UML, a model driven approach supported by aspect-oriented programming in an attempt to model various \gls{Crosscutting} concerns of context-aware applications at an early stage of the software development. Theme/UML provides a systematic means to analyse the requirements' specification in order to identify base and \gls{Crosscutting} concerns, and the relationships between them. However, to the best of our knowledge, there is no similar approach that can help the developers to analyse and understand the context-dependent behaviours in the requirements, design and implementation of the self-adaptive applications. 

The Theme/UML approach was based on the use of the \gls{MOF} extension and the \gls{ECORE} \cite{EMF:2010}. The \gls{MOF} meta model for the development of context-aware mobile applications proposed by de Farias et al. \cite{deFarias:2007p1244209} was structured according to the core and service views of the software system. This approach provides a contextual model that is independent from the application domain. However, it does not provide high-level abstraction of the software models, which express conceptual characteristics of the context-dependent behaviours. From a software developer's perspective, it does not take into account architectural or deployment issues, because it is based on the service-oriented architectures. In addition, it has focused on the model-to-model transformation for generating the software composition. Such approach adds substantial overhead over the development for writing and configuring the \gls{MOF} scripts. The Theme/UML methodology limits the development of self-adaptive applications to a very specific framework that supports the \gls{AspectJ} and \gls{EMF} \cite{Kiczales:2001p3222}. Extending this paradigm for another platform requires a specific compiler that supports \gls{AOP} and toolset that follow the \gls{EMF}.

Plastic is another development approach, which uses the MDD paradigm for developing and deploying adaptable applications, implemented in Java language \cite{Inverardi:2009p2345}. The Plastic development process focuses on the model verification and validation and service composition of java service stubs. The methodology shows a very interesting feature of runtime model verification and validation mechanism. Unfortunately, the generated software is tightly coupled with the target deployment platform and cannot be used with a standard development process supported by a standard object-oriented language other than the JAVA and \gls{AspectJ} languages. However, the two-paradigm Theme/UML and Plastic face challenges with regard to the model manipulation and management. These challenges arise from problems associated with (1) defining, analysing, and using model transformations, (2) maintaining traceability links between model elements to support model evolution and round-trip engineering, (3) maintaining consistency among viewpoints, (4) tracking versions, and (5) using models during runtime \cite{France:2007p54709}.

\subsection{A-MUSE }
An MDA-based approach for behaviour modelling and refinement is introduced by Daniele et al. \cite{Daniele:2009p3511}. Daniele et al. proposed the \gls{A-MUSE} approach, which focuses on the decomposition of the \gls{PIM} model into three levels; each level is used to automate a behavioural model transformation process. Daniele et al. \cite{Daniele:2009p3511} applied their approach to a Mobile System \gls{DSL} (called M-MUSE). Therefore, the platform independent design phase has been decomposed into the service specification and platform-independent service design steps. The platform-independent service design model should be a refinement of the service specification, which implies correctness and consistency, particularly of behavioural issues, which have to be addressed in the refinement transformation. However, when trying to realize this refinement transformation, a gap between service specification and platform-independent service design was wide, so that correctness and consistency were hard to guarantee in a single refinement transformation. The authors approach this problem by proposing multiple rounds of transformation between the \gls{PIM} and \gls{PSM}, which requires the developers to switch simultaneously between the \gls{PIM}, \gls{PSM} and the service specifications several times.

 \subsection{CAMEL}

\gls{CAMEL} is an MDD-based approach proposed by Sindico and Grassi \cite{Sindico:2009p3478}. The approach uses a domain-specific language called \gls{JCOOL}, which provides a metamodel for context sensing with the supports of the context model designed using the \gls{JCOOL} meta model. However, Sindico and Grassi implemented the context binding as the associate relationship between context value and context entity. On the other hand, context-driven adaptation refers to a structure or behaviour elements, which are able to modify the behaviour based on context values. The structural or behavioural insertion is accomplished whenever a context value changes; it uses \gls{AOP} inter-type deceleration, where the behavioural insertion is accomplished by means of an \gls{AOP} \gls{Advice} method to inject a specific code into a specific \gls{Joinpoint}.  

The \gls{CAMEL} paradigm provides insufficient details with regard to the underlying component model or the application architecture. The authors used their former domain-specific language to support the \gls{COP} approach proposed by Hirschfeld et al. \cite{Hirschfeld:2008p1620}. Moreover, \gls{CAMEL} has no formal MDD methodology that possesses a generic life cycle that a developer can use. Irrespective of these problems, \gls{JCOOL} is specific to an \gls{AOP} framework called the Simple Middleware Independent LayEr (SMILE) \cite{Smile:4432}. SMILE platform used for distributed mobile applications \cite{Smile:4432}. The model approach in \gls{JCOOL} supports only ContextJ, which is an extension of the Java language proposed by Appeltauer et al. \cite{Appeltauer:2009p3541}. The \gls{CAMEL} methodology requires the software to be re-engineered whenever a new context provider is introduced into the context model. The developers must build a complete context model for the new values and maintain the underlying \gls{JCOOL} \gls{DSL} and the \gls{UML} model. The \gls{CAMEL} methodology has adapted \gls{AOP} and the \gls{EMF} to produce a context-oriented software similar to the layered approach proposed by Hirschfeld et al. \cite{Hirschfeld:2008p1620}. This makes \gls{CAMEL} limited to the \gls{EMF} tool support and the ContextJ language \cite{Haupt:2010p3399}. From our point of view \gls{CAMEL} tightly coupled the software with modelling language, modelling tool and the target deployment platform. 

 
\subsection{MUSIC MDD}

The \gls{MUSIC} development methodology \cite{MUSIC:2009p3577} adapts a model-driven approach to construct the application variability model. In \gls{MUSIC}, applications are built using a component framework, with component types as variation points. The MUSIC middleware is used to resolve the variation points, which involves the election of a concrete component as a realization for the component type. The variability model defines the component types involved in the application's architecture and describes their different realizations. This comprises either a description of collaborating component types and rules for a composite realization, or a reference to a concrete component for an atomic realization. To allow the realization of a component type using external services, the variability model also includes a service description, which is used for service discovery. 

The software architecture in \gls{MUSIC} is a pluggable architecture for self-adaptive applications. It proposes middleware featuring a generic and reusable context management system. The architecture supports context variation and resource utilization by separating low-level platform-specific context from higher-level application-specific concerns. The resource utilization is improved through intelligent activation and deactivation of context-related plug-ins based on the needs of the active application. The \gls{MUSIC} middleware architecture defines multiple components that interact with each other to seamlessly enable self-adaptive behaviour in the deployed applications. These components include context management, adaptation reasoner, and a plug-in life-cycle management based on the \gls{OSGI} \cite{OSGI:2010p1}.

 At runtime, a utility function is used to select the best application variant; this is the so-called 'adaptation plan'. The utility function is defined as the weighted sum of the different objectives based on user preferences and QoS. Realistically, it is impossible for the developer to predict all possible variations of the application when unanticipated conditions could arise. In addition, mobile computing devices have limited resources for evaluating the many application variations at runtime and can consume significant amounts of device resources. As an outcome, the benefit gained from the adaptation is negated by the overhead required to achieve the adaptation \cite{Salehie:2009p3693}.

 \subsection{Paspallis MDD}
Paspallis \cite{Paspallis:2009p3397} introduced a middleware-centric development of context-aware applications with reusable components. Essentially, his work is based on the \gls{MUSIC} platform \cite{Reichle:2008p2719}. According to Paspallis, an MDA-based context-aware application is built by separating the concerns of the context provider from those of the context consumer. For each context provider, a plug-in or bundle is planned and designed during the design phase. At runtime, a utility function is used to consider the context state and perform decision-making process. Once the plug-in is selected to be loaded into the application, middleware support performs dynamic runtime loading of the plug-in.

 However, it is impossible for the developers to predict all the context providers that might produce context information at runtime. In addition, using this methodology means that the developer is required to design a separate plug-in architecture for each context provider, which is proportional to the available number of context providers. Additionally, this methodology does increase the development effort as each plug-in requires a separate development process.

 \subsection{U-Music MDD}
Khan \cite{Khan:2010p4084} proposed \gls{U-MUSIC} methodology. \gls{U-MUSIC} adapts a model-driven approach to constructing self-adaptive applications and enabling component model-based, unanticipated adaptation. However, the author has modified the MUSIC methodology to support semi-anticipated adaptation; also called planning-based adaptation, which enables the software to adapt among foreseeable context changes. \gls{U-MUSIC} enables developers to specify the application variability model, context elements, and data structure. The developers are able to model the component functionalities and quality of service (QoS) properties in an abstract, platform-independent way. In \gls{U-MUSIC}, dynamic decision-making is supported by the MUSIC middleware mentioned above. However, this approach suffers from a number of drawbacks. First, it is well-known that correct identification of the weight for each goal is a major difficulty for the utility function. Second, the approach hides conflicts between multiple goals in its single, aggregate objective function, rather than exposing the conflicts and reasoning about them. It would be optimistic to assert that the process of code generation from the variability models can become completely automatic or that the developer's role lies only in application design. 
\subsection{CAUCE}

\gls{CAUCE} proposed as a model-driven development approach \cite{Tesoriero:2010p927367}. The authors defined an MDA approach that focuses on three layers of models. The first layer confirms to the computational independent model for capturing the conceptual properties of the applications. The second layer defines three complementary points of view of the software systems. These views include deployment, architecture and communication. The third layer focuses on converting the conceptual representation of the context-aware application into a software representation using a multi model transformation. The \gls{ATL} is used to interpret the model and convert them into a set of models conforming to the platform independent model. The final model is transformed using the \gls{MOF} Script language based on the \gls{EMF} paradigm \cite{EMF:2010}. The \gls{CAUCE} methodology focuses more on the \gls{CIM} by splitting this layer into three layers of abstraction, which confirms to the tasks, social and space meta models. The task model focuses on modelling a set of tasks and the relationships among them that any entity in the system is able to perform. The social metamodel defines the social environment of the entities in the system and is directly related to the entity task and entity information that identify of the context-aware application behaviour. The space metamodel defines the physical environment of the entities in the system. Therefore, this metamodel is directly related to the physical conditions, infrastructure and location characteristics of the context-aware applications.

However, the \gls{CAUCE} methodology provides a complete development process for building context-aware applications. Despite that, \gls{CAUCE} is limited to specific modelling tool and language, in this case the \gls{UML} is integrated with \gls{EMF}. The generated application can only be implemented using Java language as it is supported by the \gls{ATL} and \gls{MOF} Script languages. However, it is impossible for the developers to adapt \gls{CAUCE} for building heterogeneous and distributed mobile applications, which might have multiple deployment platforms and requires variant implementation languages.  
\subsection{ContextUML}
Generally, \gls{UML} profiles and metamodels are used to extend the \gls{UML} language semantics. ContextUML was one of the first approaches that targeted the modelling of the interaction between context and web service applications \cite{Sheng:2003p3501}. ContextUML was extended by Prezerakos et al. \cite{Prezerakos:2007p3774}, using aspect-oriented programming and service-oriented architecture to fulfil the user's needs. However, contextUML used a \gls{UML} metamodel that extended the regular \gls{UML} by introducing appropriate artifacts that used to create context-aware applications. contextUML  produces a class diagram, which corresponds to the context class and to specific services. They mitigate the \gls{UML} relationship and dependency to express the interaction between the context information and the respective services. A means of parameter injection and service manipulation are used to populate specific context-related parameters in the application execution loop.

However, the \gls{UML} profiles and metamodels lack from several features required for modelling the self-adaptive software system. Ignoring the heterogeneity of the context information, they based their claims on the nature of the context values, which can fluctuate and evolve significantly at runtime. It is not feasible to this study how the behaviour is modelled when multiple context values have changed at the same time. 
\subsection{COCA-MDA}

Accodring to the \gls{COCA-MDA} approach, the software self-adaptability and dependability can be achieved by dynamically composing software from context-oriented modules based on the context changes rather than composing the software from functional-oriented modules. Such composition requires the design of software modules to be more oriented towards the context information rather than being oriented towards the functionality. The principle of context orientation of software modules was proposed in the \gls{COCA-MDA}  \cite{magableh:2011p1231}. \gls{COCA-MDA} proposes a decomposition mechanism of software based on the separation between context-dependent and context-independent functionality. Separating the context-dependent functionality from the context-independent functionality enables adaptability and dependability of software systems with the aid of middleware technology. The middleware can adapt the software behaviour dynamically by composing interchangeable context-dependent modules based on context changes. \gls{COCA-MDA} proposes that software self-adaptability is achieved by having both an adaptive middleware architecture and a suitable decomposition strategy, which separates the context-dependent functionality from the context-independent functionality of the software systems. 

\gls{COCA-MDA} was proposed as a generic and standard development paradigm towards constructing self-adaptive software from context-oriented components, which enables a complete runtime composition of the context-dependent behaviours and provides the software with capabilities of self-adaptability and dependability in mobile computing environment. The context-oriented component model encapsulates the implementation of the context-dependent parts in distinct architectural units, which enables the software to adjust its functionality and/or behaviour dynamically. This differs from the majority of existing work, which seek to embed awareness of context in the functional implementation of applications. The context-oriented software is developed using a \gls{COCA-MDA}. Afterwards, the context-oriented software is manipulated at runtime by a \gls{COCA-middleware} that performs a runtime behavioural composition of the context-dependent functionality based on the operational context. The self-adaptive software dependability is achieved through the \gls{COCA-middleware} capability in considering its own functionality and the adaptation impact/costs. A dynamic decision-making based on a policy framework is used to evaluate the architecture evolution and verifies the fitness of the adaptation output with the application's objectives, goals and the architecture quality attributes. 

The \gls{COCA-MDA} follows the principles of \gls{OMG} model-driven architecture. In \gls{MDA}, there are three different viewpoints of the software: the \gls{CIV}, the \gls{PIV}, and the \gls{PSV}. The \gls{CIV} focuses on the environment of the system and the requirements for the system, and hides the details of the software structure and processing. The \gls{PIV} focuses on the operation of a system and hides the details that are dependent on the deployment platform. The \gls{PSV} combines the \gls{CIV} and \gls{PIV} with an additional focus on the details of the use of a specific platform by a software system \cite{OMG:2010p3514}. COCA-MDA partitioning the software into three viewpoints: the structure, behaviour, and enterprise viewpoints. The structure viewpoint focuses on the core component of the self-adaptive application and hides the context-driven component. The behaviour viewpoint focuses on modelling the context-driven behaviour of the component, which may be invoked in the application execution at runtime. The enterprise viewpoint focuses on remote components or services, which may be invoked from the distributed environment.  The \gls{COCA-MDA} provides the developers with the ability to specify the adaptation goals, actions, and causes associated with several context conditions using a policy-based framework. For each \gls{COCA-component}, the developers can embed one or more \glspl{DPL} that specify the architecture properties. The \gls{DPL} is described by a state-machine model based on a set of internal and external variables and conditional rules. The rules determine the true action or else an action based on the variable values. The action part of the state diagrams usually involves invoking one or more of the component's layers. A single layer is activated if a specific context condition is found, or deactivated if the condition is not found.

 
 The use of COCA-MDA for building self-adaptive applications for indoor wayfinding for individuals with cognitive impairments was proposed in \cite{magableh:2011p0002}. Evaluating the COCA-MDA productivity among the development cost and effort using \gls{COCOMO II} \cite{Boehm:2008p4159} was demonstrated in \cite{magableh:2011p1112} . This article focuses on evaluating the features of COCA-MDA against the above mentioned approaches as shown in the following section.

 \section{Feature Analysis and Comparative Study of MDA-based Approaches\label{sec:feature}}

From the software developer's perspective, it is vital to know the features of the development paradigm, which might be used in constructing a self-adaptive application. Feature evaluation of the development methodology can assist the developers in selecting among the proposed methodologies in the literature for achieving adaptability and dependability of the software systems. Improving the development of self-adaptive software systems using model driven approach has attained several research efforts. The target was in general to introduce software with adaptability and variability while focusing on reducing the software complexity and optimising the development effort. 

The examination of software system performance, dependability and availability is of greatest importance for tuning software system in conjunction with several architecture quality attributes. Such performance analysis was considered by the \gls{MOSEL} \cite{Begain:2001p558125}. The system modelling using \gls{MOSEL} illustrates how easily it can be used for modelling real-life examples from the fields of computer communication and manufacturing systems. However, extending the \gls{MOSEL} language towards the modelling and performance evaluation of self-adaptive software system can estimate several quality attributes of model-based architecture and provides early results about how efficient is the adaptation action.  

Kitchenham et al. in \cite{Kitchenham:2002p3826} proposed the \GLS{DESMET} method, which evaluates software development methodologies using an analytical approach. Asadi et al. have adapted the \GLS{DESMET} method to analyse several MDA-approaches. The authors adapted several evaluation criteria that can be used to compare \gls{MDA} methodologies based on MDA-related features and MDA-based tool features \cite{Asadi:2008p3532}.

However, Calic et al. \cite{Calic:2008p3829} proposed an evaluation framework to evaluate MDA-based approaches in terms of four major criteria groups, as follows: I) MDA-related features: The degree to which the proposed methodologies are compliant with OMG's MDA specification \cite{OMG:2010p3514}. II) Quality: Evaluation of the overall quality of the MDA-based approaches including their efficiency, robustness, understandability, ease of implementation, completeness, and ability to produce the expected results \cite{Kitchenham:2002p3826}. III) Usability: Simplicity of use and ease of implementation by the developer, which covers clear information about the impact of the methodology on the development effort \cite{ANorman:2002p3830,sharp:2007p3735}. IV) Productivity: The quality of benefits derived from using the methodology and its impact on the development time, complexity of implementation, code quality, and cost effectiveness \cite{Calic:2008p3829}. Calic at al. \cite{Calic:2008p3829} presents the COPE tool, to evaluate the MDA productivity by automate the coupled evaluation of metamodels and model by recording the coupling history in an history model.

Lewis et al. \cite{Lewis:2005p4171} have evaluated the impact of MDA on the development effort and the learning curve of the MDA-based development tools based on their own experiences. The authors concluded that the real potential behind MDA is not completely employed either by current tools or by the proposed MDA approaches in the literature. In addition, the developers have to modify the generated code such that it is suitable for the target platform. The MDA tools can affect the level of maintenances required for the generated codes. In the same way, the developer's level of understanding  of MDA tasks and  their familiarity with the target platform have direct impacts on MDA productivity. 

The \gls{COCOMO II} \cite{Boehm:2008p4159} emerged as software cost estimation model, which considers the development methodology productivity. The productivity evaluates the quality of benefits derived from using the development methodology, in terms of its impact on the development time, complexity of implementation, code quality, and cost effectiveness \cite{Calic:2008p3829}. \gls{COCOMO II} allows estimation of the effort, time, and cost required for software development. The main advantage of this model over its counterparts such as the Software Life-cycle Management (SLIM) model \cite{Estell:1976p3432} and the System Evaluation and Estimation of Resources Software Estimation Model (SEER-SEM) \cite{Galorath:2006p46325} is that \gls{COCOMO II} is an open model with various parameters which effect the estimation of the development effort. Moreover, the \gls{COCOMO II} model allows estimation of the development effort in \gls{PM} and the \gls{TDEV} a software application. A set of inputs such as software scale factors (SF) and 17 effort multipliers is needed. A full description of these parameter is given in the \gls{COCOMO II} model definition manual, which can be found in \cite{Boehm:2008p4159}. An example of an evaluation of MDA approaches with \gls{COCOMO II} can be found in \cite{Achilleas:2010p4154}.

In this research, we intend to use an evaluation framework that can test and qualify the ability of MDA-based approaches to produce the expected results \cite{Kitchenham:2002p3826} in terms of dynamic adaptation in general, and self-adaptability, in specific. These features are evaluated in the following sections. 

\subsection{Existence of MDA-related Features}

MDA features refers to the degree to which the proposed methodologies are compliant with the OMG's MDA specifications; these specifications can be divided into the support of \gls{CIM}, \gls{PIM}, \gls{PSM}, model validation, and transformation \cite{OMG:2010p3514}. In terms of MDA features, we adapt the criteria proposed by Asadi and Ramsin \cite{Asadi:2008p3532}, which highlights the methodology's conformance to the original OMG standard, as shown in Table \ref{table_MDAFeatures}. Feature analysis can be performed in two ways: scale form and narrative form. The scale form attaches the methodology complaint to a specific feature, which is divided into three ranks, from A to C, as shown in each table. The narrative form captures whether the methodology covers a specific feature based on the level of involvement.

 \begin{table*}[!ht]
 \centering
\caption{MDA-related criteria evaluation}
\includegraphics[scale=0.40]{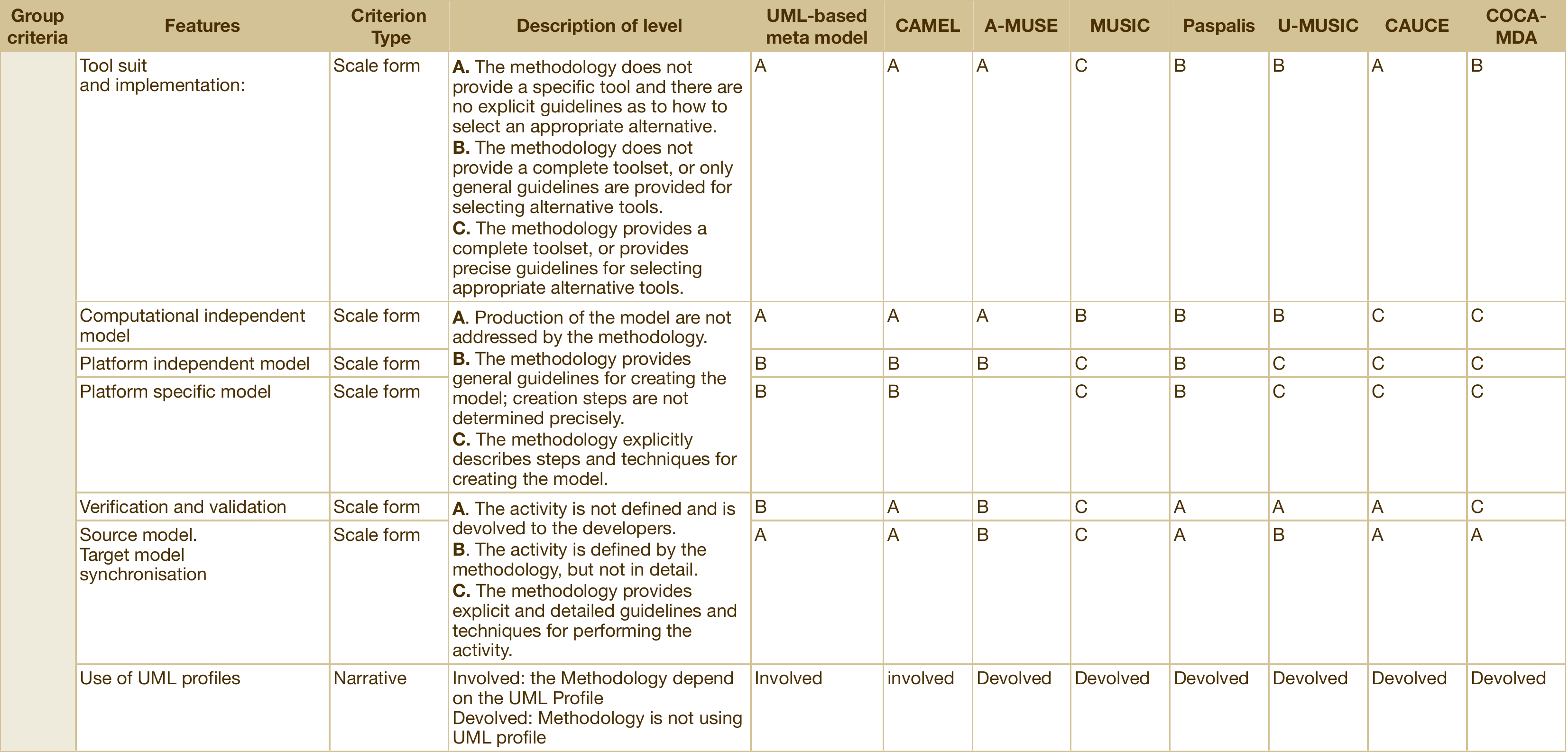}
\label{table_MDAFeatures}
\end{table*}

\subsection{Tool-related feature analysis}
The major challenges that developers face when attempting to realize the MDD vision is the selection of a modelling language and modelling tool. Modelling languages challenges arise from concerns associated with providing support for creating and using an appropriate modelling abstraction for analysing and designing the software \cite{France:2007p54709}. A second challenge posed by Asadi and Ramsin \cite{Asadi:2008p3532}, that each development methodology generates more specific technical details that suit the underlying modelling language or modelling tool they used, as each tool requires a learning curve, and it might have some limitation with regard to the platform and the number of implementation languages they support \cite{Lewis:2005p4171}. This implies that a MDD approach should be decoupled from using a specific tool or modelling language. The developers have to be free on selecting the tool(s) that fits their needs and the software under development. 

On the other hand, MDD approaches should focus more on describing standard development processes without relaying on a specific technology or platforms like \gls{EMF} and \gls{ECORE}. In terms of the tools the methodology used, the features that highlight the methodology dependency on the modelling languages and tools are shown in Table \ref{table_MDAFeaturesanalysis}.  

\begin{table*}[!ht]
\centering
\caption{MDA tool-related criteria}
\includegraphics[scale=0.46]{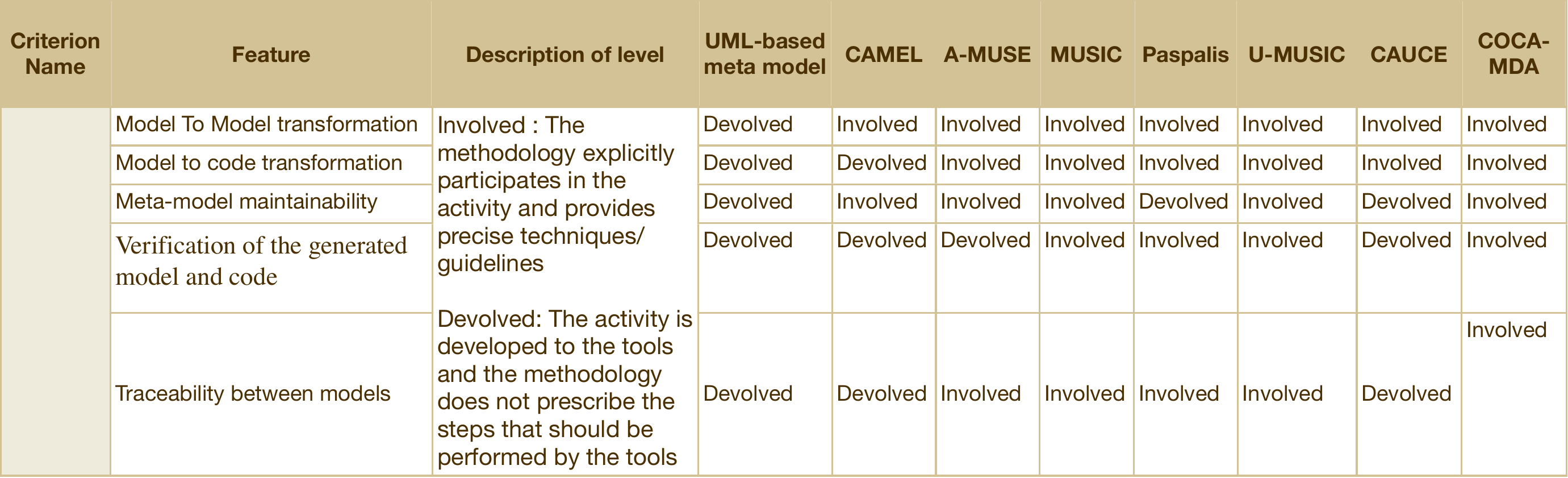}
\label{table_MDAFeaturesanalysis}
\end{table*}

\subsection{Quality of the MDA-based Approaches}

 Quality refers to the overall quality of the MDA-based approaches, including their efficiency, robustness, understandability, ease of implementation, completeness, and ability to produce the expected results \cite{Kitchenham:2002p3826}. However, in this research, we have focused on the ability of the MDA-based approaches to provide the expected results that support the adaptability of the generated software, whether these results are derived from the code or the architecture. Moreover, we have split these criteria into four groups: requirements engineering, unanticipated awareness, context model, and modelling context-dependent behavioural variations.

\subsubsection{Requirements Engineering of Context-dependent Behavioural Variations} 

Requirements engineering refers to the causes of adaptation, other than the functional behaviour of the self-adaptive system. Whenever the system captures a change in the context, it has to decide whether it needs to adapt. The MDA-based approaches in the related work were evaluated regarding whether they support the modelling of context requirements as a specific feature and whether they support the requirements' engineering in general, as shown in Table \ref{table_reqref}. 

In addition, the methodology's ability to analyse and models the context-dependent behaviour variations requires the MDD supports at three levels. The first is the requirement analysis at the computational independent model. The second is the representation of these requirements by means of \gls{UML} objects at the platform independent model and platform specific model. The third is the representation of the context-dependent behaviour as runtime objects, which are a code representation of these requirements \cite{Bencomo:2010p3675}. However, the evaluation of these criteria is shown in Table \ref{table_reqref}.  

\begin{table*}[!ht]

\caption{Supporting context-dependent behaviour variations on the analysis, design and implementation}
\centering
\includegraphics[scale=0.45]{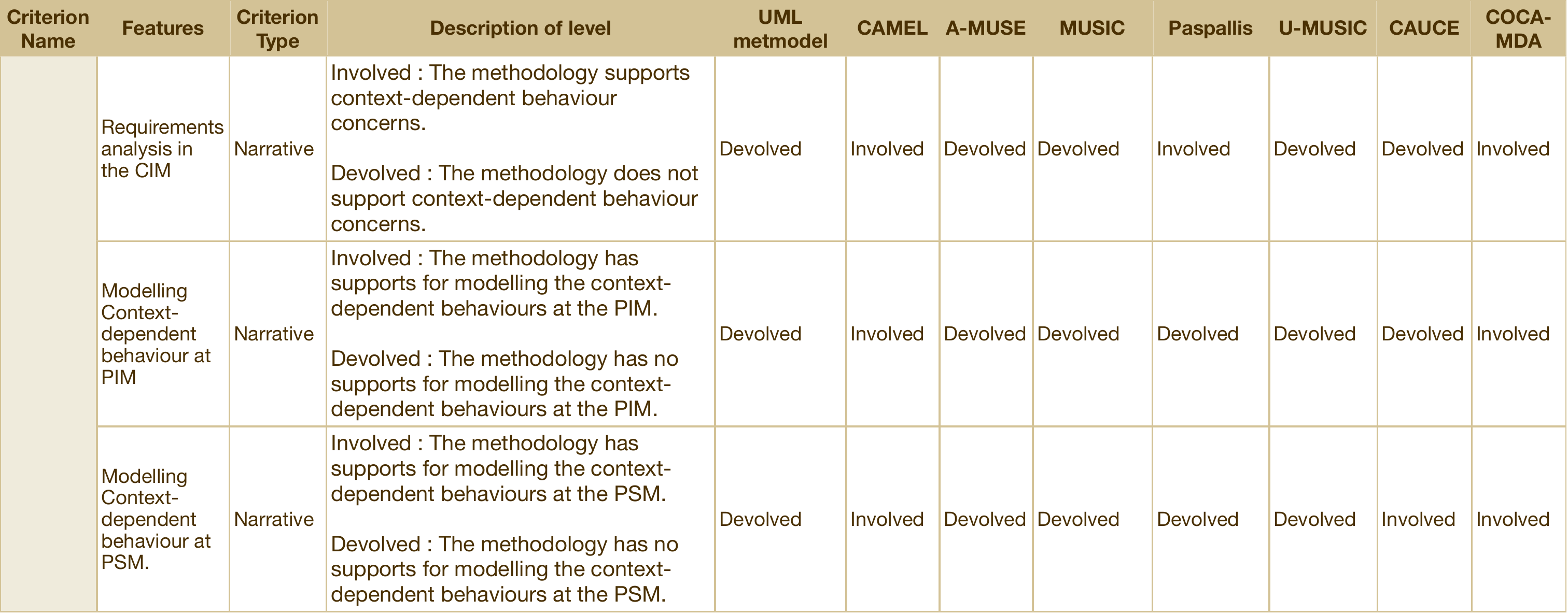}
\label{table_reqref}
\end{table*}

\subsubsection{Unanticipated Awareness}

This feature captures whether a context change can be predicted ahead of time \cite{Cheng:2009p3763}. Anticipation can be classified into three degrees: foreseen, foreseeable, and unforeseen changes. Foreseen refers to the changes that are handled in the implementation code. Foreseeable refer to the context changes that were predicted at the software design. Unforeseen refers to the changes that are not modelled at the design or the implementation stage, but are to be handled at runtime \cite{Laprie:2008p3772}. The evaluation criteria are shown in Table \ref{table_anticpation} with their related scale form.

\begin{table*}[!ht]
\centering
\caption{Anticipation of context change-related criteria and evaluation results}
\includegraphics[scale=0.43]{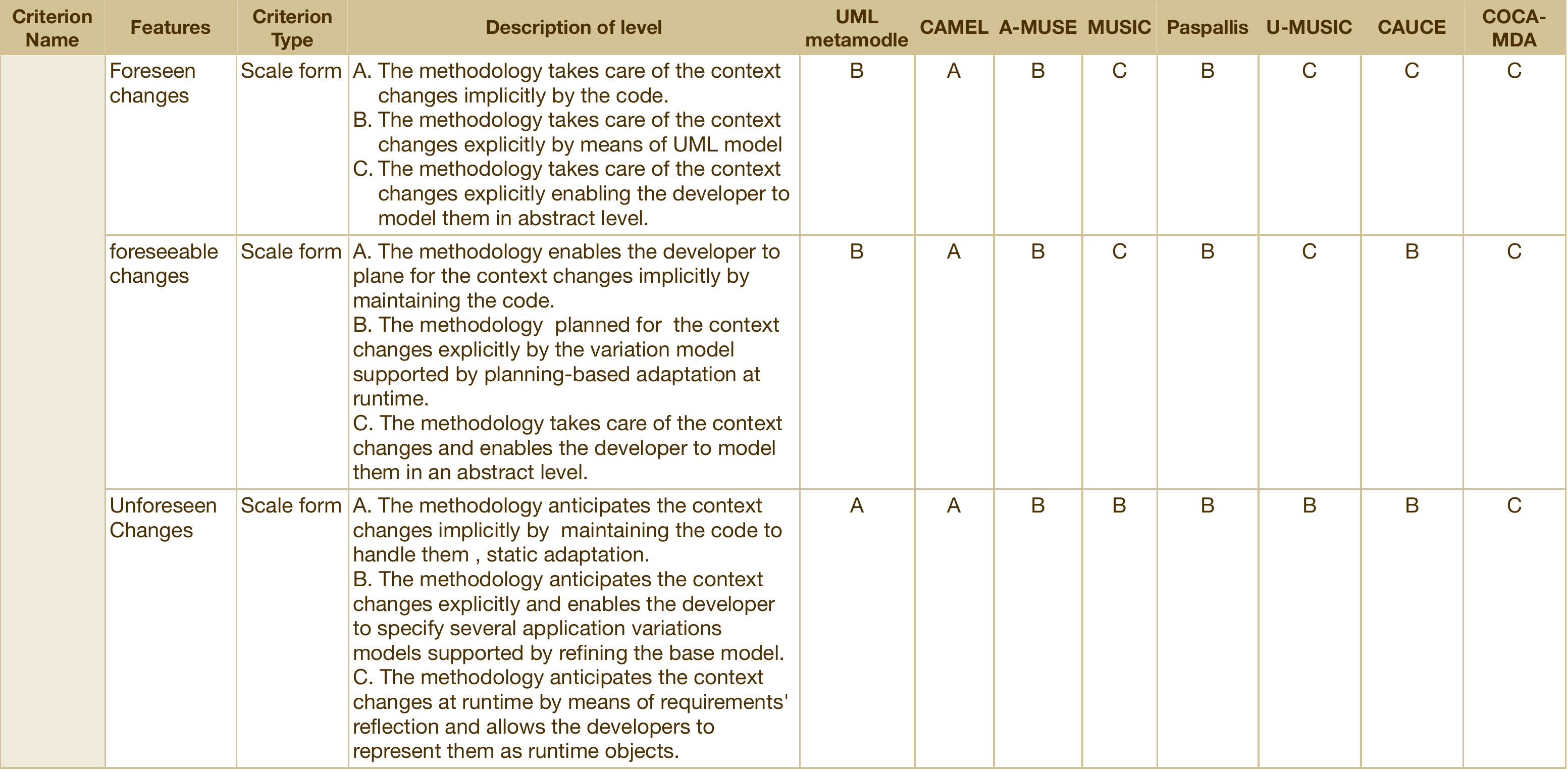}
\label{table_anticpation}
\end{table*}

\subsubsection{Context Model}
This captures the ability of the methodology to incorporate the context information using the '\gls{Separation of concerns}' technique between the context information model and the business logic. The first criterion focuses whether the methodology supports/uses the \gls{Separation of concerns} in the development processes. The second criterion refers to the ability to bind the context source to the context provider, as proposed by Sen and Roman \cite{Sen:2010p3766} and Broens et al. \cite{Broens:2007p3698} and Paspallis \cite{Paspallis:2010p574}. The binding mechanism enables the developers to map each context cause to the affected architectural units. The binding mechanism also enables the application to determine which part has to manage the context changes, by means of the adaptation mechanism. 

The evaluation criteria for the context model are shown in Table \ref{table_contextmodel}.
\begin{table*}[!ht]
\caption{Context model-related criteria and evaluation results}
\centering
\includegraphics[scale=0.45]{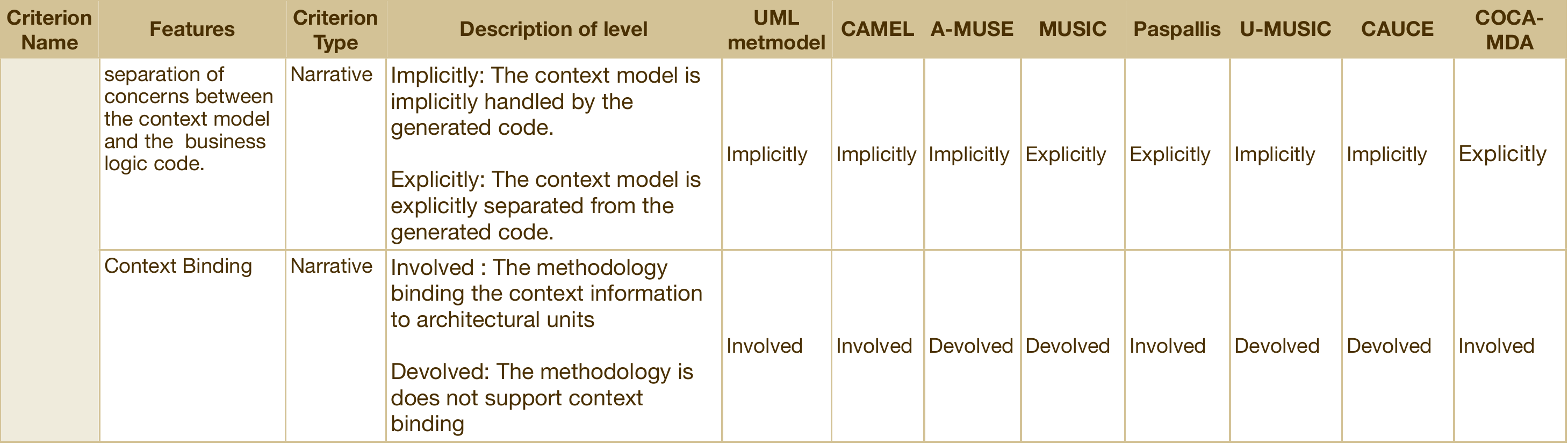}
\label{table_contextmodel}
\end{table*}
\subsubsection{Modelling Context-dependent Behaviour}
 These criteria refer to the ability of the model to capture the impact of context changes on the self-adaptive application's behaviour. However, Hirschfeld et al. \cite{Hirschfeld:2008p1620} classified these changes into three kinds of variations: actor dependent, system dependent, and environment dependent behavioural variations. These behavioural variations requires a separation between their concerns, by separating the context handling from the concern of the application business logic.
In addition, a separation between the application-dependent parts from the application-independent parts can support behavioural modularization of the application, thereby simplifying the selection of the appropriate parts to be invoked in the execution, whenever a specific context condition is found. The behavioural modelling criteria are shown in Table \ref{table_modelbehavioir}.


\begin{table*}[!ht]
\caption{Modelling context-dependent behaviour variations and evaluation results}

\centering
\includegraphics[scale=0.45]{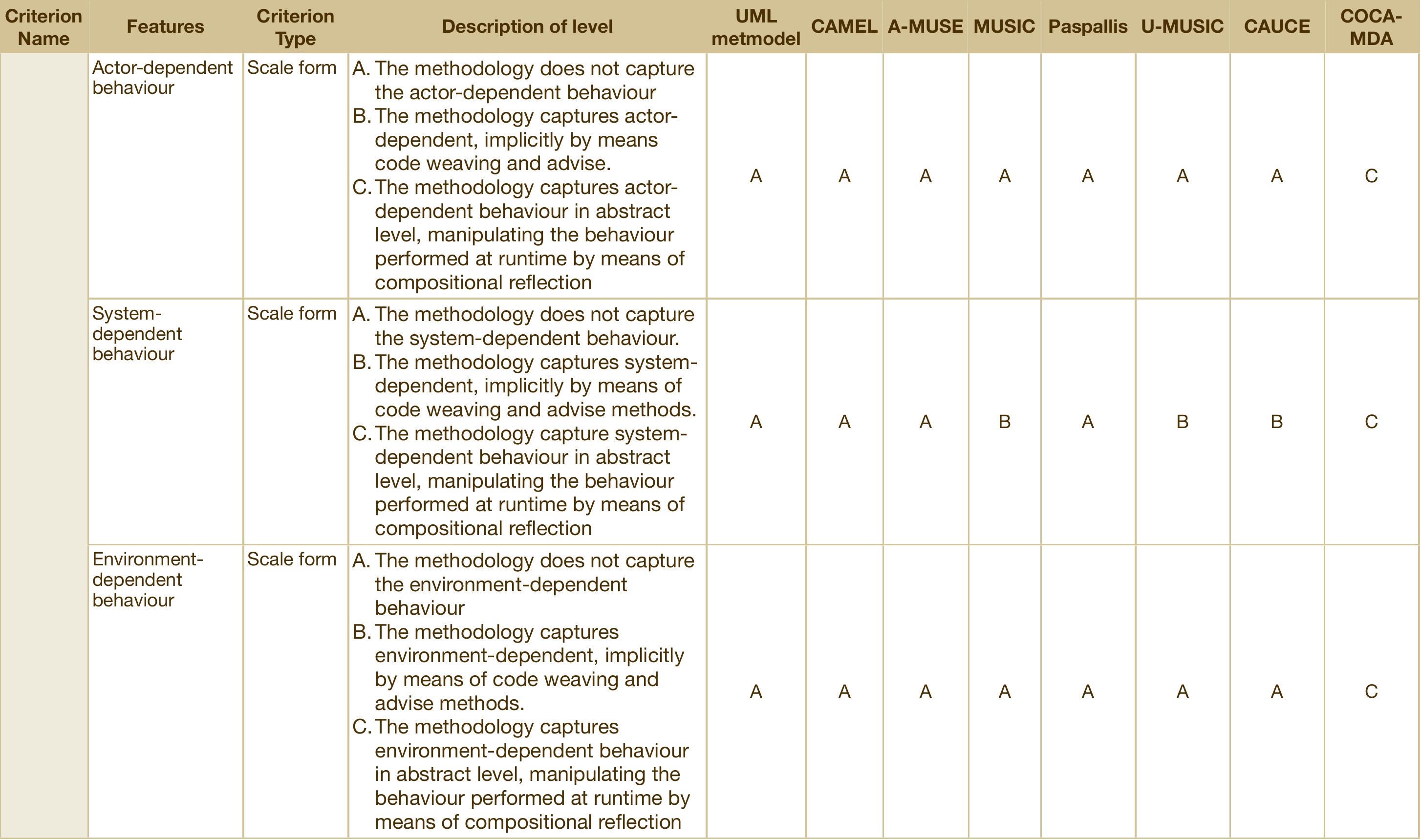}
\label{table_modelbehavioir}
\end{table*}

\section{Evaluation Results \label{sec:res}}
Based on the analysis results shown in Tables 
\ref{table_MDAFeatures}, \ref{table_MDAFeaturesanalysis}, \ref{table_reqref}, 
 \ref{table_anticpation}, \ref{table_contextmodel}, and \ref{table_modelbehavioir},
we find that the discussed methodologies in the related work suffer from several critical failings in terms of their conformance to the \gls{OMG}'s guidelines for \gls{MDA} methodology \cite{OMG:2010p3514}. 

First, it is well known that correct identification of the weight of each goal is a major difficulty for the utility functions as shown in the \gls{MUSIC}, \gls{U-MUSIC} and Paspallis methodologies.

 Second, these approaches hide conflicts among multiple adaptation goals by combining them into a single, aggregate objective function, rather than exposing the conflicts and reasoning about them. On the other hand, it would be optimistic to assert that the process of code generation from models can become completely automatic or that the developer's role lies only in application design, as discussed in the above with regard to \gls{CAMEL} and \gls{A-MUSE}. 
 
 Third, it is impossible for the developers to predict the possible application variations, which will extend the application behaviour when unanticipated conditions arise, this applied to all methodologies mentioned in the above.

 In addition, mobile devices have limited resources for evaluating many application variations at runtime, which might consumes significant amounts of the allocated resources. As a result, the benefits gained from the adaptation are negated by the overhead required to achieve the adaptation. Fourth, the previously mentioned methodologies produce an architecture with a tight coupling between the context provider and the context consumer, which may cause the middleware to notify multiple components about multiple context changes. Finally, all the methodologies seem to generate an architecture that is tightly coupled with the target platform for deployment and the modelling tools they used.  

In addition, the developers have to explicitly predict the final composition of the software and the possible variations of the application, whether at the platform independent model or through the model transformation. Moreover, the developers have to modify the generated code to be suitable for deployment on the target platform and to be integrated with the middleware implementation, which is in the best case made a hug gap between the middleware designer and the application developer.  Understandings the modelling tasks and the target platform configurations have limited software developers from employing MDA-approaches in several platforms.

\section{Conclusions }
With the current state-of-the-art, it is possible to design a system that could adapt its behaviour. However, any adaptation would either have to be pre-defined at design time, or would have to be a reflective response to some monitored parameters, perhaps by using the available techniques of model-driven architecture. An effective pre-defined responses would be dependent on the requirements analyst anticipating and enumerating of all the possible environmental states and the corresponding behaviour required. A drawback of the reflective response is that the relationship between the adaptation and the objective goal would be at best implicit, making the verification of goal satisfaction hard or even impossible. The MDD-based approaches proposed in the literature suffer from a number of drawbacks. First, it would be optimistic to assert that the process of model transformation and code generation from the software models can become completely automatic and that the developer's role lies only in application design. Second, it is impossible for the developer to predict all possible variations of the application when unanticipated conditions will arise. In addition, mobile devices have limited resources for evaluating many application variations at runtime and can consume significant amounts of device resources. As result, the benefit gained from the adaptation is negated by the overhead required to achieve the adaptation. Third, each development methodology generates more specific technical details that suit the underlying implementation language or modelling tool they used. We found that \gls{COCA-MDA} is more capable to meet the requirements of self-adaptive systems and allows the software developers to adapt it in wide range of execution platforms.

\bibliographystyle{splncs}
\bibliography{./referenecev11-5-2011.bib}

\balance
  \end{document}